\documentclass[12pt]{article}

\usepackage{amssymb}
\usepackage{amsmath}
\usepackage{bm}

\def\iu{i_U}

\def\dualu{{\dual{U}}}

\def\man{{M}}

\def\dual#1{{\widetilde{#1}}}
\def\GamTM{{\Gamma T\man}}
\def\GamLamM#1{{\Gamma \Lambda^{#1}\man}}

\def\qquadand{\qquad\text{and}\qquad}

\newcommand\BE[1]{{\begin{equation}#1\end{equation}}}
\newcommand\BAE[1]{{\begin{equation}{\begin{aligned}#1\end{aligned}}\end{equation}}}

\def\iuF{\iu F}
\def\iustarF{\iu{{\star}}F}

\newcommand{\bfe}{ \TYPE 1    {\mathbf e}  {}   }
\newcommand{\bfb}{ \TYPE 1 {\mathbf b}   {}}
\newcommand{\bfd}{ \TYPE 1 {\mathbf d}   {} }
\newcommand{\bfh}{\TYPE 1  {\mathbf h}   {} }

\newcommand{\bfE}{ \TYPE 2    {\mathbf E}  {}   }
\newcommand{\bfB}{ \TYPE 2 {\mathbf B}   {}}
\newcommand{\bfD}{ \TYPE 2 {\mathbf D}   {} }
\newcommand{\bfH}{\TYPE 2  {\mathbf H}   {} }

\newcommand\bbfJ{{\bf J}}

\newcommand\bbfj{{\bf j}}

\newcommand{\bfEdot}{ \TYPE 2    {\dot{\mathbf E}}  {}   }
\newcommand{\bfBdot}{ \TYPE 2 {\dot{\mathbf B}}   {}}
\newcommand{\bfDdot}{ \TYPE 2 {\dot{\mathbf D}}   {} }
\newcommand{\bfHdot}{\TYPE 2  {\dot{\mathbf H}}   {} }

\def\Me{{\mathbf e}}
\def\Mb{{\mathbf b}}
\def\Md{{\mathbf d}}
\def\Mh{{\mathbf h}}

\newcommand\TYPE[3]{ \underset {(#1)}{\overset{{#3}}{#2}}  }

\newcommand\EP{\epsilon_r\epsilon_0\,}

\newcommand\MU{\mu_r\mu_0\,}

\newcommand\ep{\epsilon}

\newcommand{\hash}{\#}

\newcommand{\D}{{\boldsymbol d}\,}

\newcommand\PD[1]{\frac{\partial}{\partial #1}}

\newcommand{\DD}{{\cal U}}

\newcommand\hp{\hat\chi}

\newcommand\union\cup

\newcommand{\beq}{\begin{equation}}
\newcommand{\beqa}{\begin{eqnarray}}
\newcommand{\eeq}{\end{equation}}
\newcommand{\eeqa}{\end{eqnarray}}

\def\ee{\,\epsilon_0\,}
\def\cc{\, c_0 \,}

\newcommand\DIMM[2]{\{#1  \}  \{ #2 \}}

\newcommand\DIM[1]{\{#1  \} }

\newcommand\LAM[2]{{\Lambda}(#1,#2)}

\newcommand\STAR[1]{\TYPE {#1} {\star} {}}
 \newcommand\xx{\mathbf x}

 \newcommand\yy{\mathbf y}

\newcommand\gm[1]{\TYPE {#1,#1}\gamma {} (X,Y)}

\newcommand\KK[1]{\TYPE {#1,#1}{K_\omega} {} (X,Y)}

\newcommand\gmp[1]{\TYPE {#1,#1}{\gamma^\prime} {} (X,Y)}

\newcommand\bff{{\mathbf f}}

\newcommand\SIG[1]{\Sigma_{{\bff}}^{#1}}

\newcommand\LERAY[1]{\TYPE {#1} {\omega_{\mathbf f}}  {}}

\newcommand\BIGLERAY[1]{\TYPE {#1} {\Omega_{\mathbf f}}  {}}

\newcommand\LERAYY[1]{\TYPE {#1} {\omega^Y_{\mathbf f}}  {}}

\newcommand\HASH{\hash}

\newcommand\HASHX{\TYPE X \hash {}}

\newcommand\HASHY{\TYPE Y \hash {}}

\newcommand\DIR{\TYPE 0 {\delta_{p_0 \times R}} {}}

\newcommand\DSIG[1]{ \TYPE 0 {    {\bm\delta}_{\SIG {#1}}      } {}  }

\newcommand\DIRAC[1]{  \TYPE 0     {{\bm\delta}_{\SIG {#1} }}  {}    }

\newcommand\DIRACC[1]{  \TYPE    0 {\delta_{\SIG {#1} \times {R}}} {}       \,\,}

\newcommand\DDIRAC[1]{{\bm\delta}_{\SIG {#1}}}

\newcommand\DIRACY[1]{   \TYPE 0 {    {\bm\delta}^Y_{\SIG {#1}}      } {}         }

\newcommand\DIRACYY[1]{   \TYPE 0 {    {\bm\delta}^{Y_1\,Y_2\ldots Y_q}_{\SIG {#1}}      } {}         }

\newcommand\XX[1]{{\bf x}_0(#1)  }

\newcommand\XXP[1]{{{\bf \dot x}}_0(#1)  }

\newcommand\JAC{{\cal Q}}

\newcommand\bfA{\mathbf A}
\newcommand\JJJ{{\bar{\cal J}}}

\newcommand\rp{{r^\prime}}
\newcommand\phip{{\phi^\prime}}
\newcommand\zp{{z^\prime}}

\newcommand\pd[1]{\frac{\partial}{\partial #1}}

\newcommand\xp{x^\prime}
\newcommand\yp{y^\prime}
\newcommand\bDD{{\partial{\cal U}}}

\newcommand\GGG{{\cal G}}

\newcommand\BRK[2]{ \left(  #1\,, #2   \right)_{\DB} }

\newcommand\RR{{\bf R}}

\newcommand\DB{{{\cal U}}}

\newcommand\epr[3]{ \,(d\,{#1}^{{#2}_1} \wedge \ldots \wedge\, d\,{#1}^{{#2}_{#3}} )}
\newcommand\pp{p^{\prime}}
\newcommand\qq{q^{\prime}}

\newcommand\poph{{\bf h}}

\newcommand\DDelta{{\mathbf \Delta}}
\newcommand\Ddelta{\underbar{$\mathbf \delta$}}

\title{Differential Form Valued Forms and Distributional Electromagnetic Sources }

\author{Robin  W Tucker\footnote{r.tucker@lancaster.ac.uk}\\ The Cockcroft
Institute and  Lancaster University}

\begin{document}

\maketitle

\begin{abstract}

Properties of a fundamental double-form of bi-degree $(p,p)$ for $p\ge 0$ are reviewed in order to establish a distributional framework for analysing equations of the form $$\Delta\TYPE p \Phi {} + \lambda^2\TYPE p \Phi {}=\TYPE p {\cal S} {}$$ where $\Delta$ is the Hodge-de Rham operator on $p-$forms $\TYPE p \Phi {}$ on ${\bf R}^3$. Particular attention is devoted to singular distributional solutions that arise when the source $\TYPE p {\cal S} {}$ is a singular $p-$form distribution.
A constructive approach to Dirac distributions on (moving) submanifolds embedded in ${\bf R}^3$ is developed in terms of (Leray) forms generated by the geometry of the embedding.
This framework offers a useful tool in electromagnetic modeling   where the possibly time dependent sources of certain physical attributes,
such as electric charge, electric current and  polarization or
magnetization, are concentrated on localized regions in space.\footnote{  {02.40.Hw} {Classical differential geometry},\,
         {03.50.De}   {Classical electromagnetism}\,{41.20.-q} {Applied classical electromagnetism}}

\end{abstract}

\vskip 0.5cm

\section{Introduction}
\label{ch0}

Mathematical distributions  \cite{schwartz} play an all pervading role in the physical sciences.
They underpin the theoretical formulation of quantum mechanics and many linear systems in classical field theory and are used to extend the solution space for linear partial
differential equations to include non-smooth fields with singularities.
They also offer a useful tool in electromagnetic modeling   where the sources of certain physical attributes,
such as electric charge, electric current, polarization or
magnetization, are concentrated on localized regions in space \cite{jackson}, \cite{bladel},  \cite{skinner}, \cite{glasser}, \cite{coil} or
spacetime. These
aspects of distributions often occur together in problems in
electromagnetic theory and this article explores this
symbiosis in the unifying language of exterior double-forms.
Such a formulation has  direct relevance to many contemporary design problems in accelerator science where one is confronted with a number of difficult issues in electromagnetic computation. The effective resolution of such problems is often contingent on reliable computationally expensive numerical analysis which can greatly benefit from reliable analytic treatments. The formulation below will be illustrated by a number of applications in applied electro-magnetics.

\section{Notation}
\label{ch1}

The formulation exploits the geometric language of exterior differential
forms \cite{rwt} since this is  ideally suited to accommodate {\it local} changes of
coordinates that can be used to simplify the description of boundary
and initial-value problems and naturally encapsulates intrinsic {\it global} properties of domains.
In the following the distinction between smooth $(C^\infty)$ forms
on some regular domain and those with possible singularities or
discontinuities will be important. Furthermore smooth forms with
compact support will play a pivotal role. These will be referred to
as {\it  test forms} \cite{gelfand}, \cite{shilov}, \cite{neto}  and distinguished below  by a superposed hat.
It will also be useful sometimes to notationally distinguish $p-$forms  on
different manifolds. Thus $\TYPE p {\alpha_M}{}$ will denote a
$p-$form \footnote{i.e. $
 \alpha_M\in {\cal S} \Lambda^p(M)$, the space
 of sections of the exterior bundle of $p-$forms over $M$} on some manifold $M$
 while $\TYPE {p,q}{\gamma} {}(X,Y)$ will denote a
double-form of bi-degree $(p,q)$ (see section \ref{ch2}~) on the product manifold $X\times
Y$. Each $n-$dimensional manifold $M$ will be assumed orientable and
endowed with a preferred $n$-form $\TYPE M {\star} {} 1$ induced from a
metric tensor $g_M$.  One then has \cite{rwt} the linear Hodge operator $\star$ that maps $p-$forms to $(n-p)$-forms on $M$.  If $g_M$ has signature $t_g$ one may write \BE{
g_M=\sum_{i=1}^n e^i_M\otimes e^j_M\,\eta_{ij}} where
$\eta_{ij}=diag(\pm 1,\pm 1,\ldots \pm 1)$ and \BE{\TYPE M {\star} {} 1=
e^1_M\wedge e^2_M\wedge\ldots \wedge e^n_M} with
$t_g=det(\eta_{ij})$ 
and $\{ e^i_M\}$ is a set of basis $1-$forms in $\Gamma T^\star(M)$.

If $\alpha$ and $\beta$ are integrable $p-$forms on $\DB\subset\RR^n$
denote \BE{\int_\DB \alpha \wedge \STAR {\RR^n} \beta} by
\BE{\left(\alpha,\beta\right)_\DB\equiv\left(\beta,\alpha\right)_\DB}
Furthermore if one of the forms has compact support on $\DB$ (and
belongs to the space of Schwartz test functions) an alternative
notation is \BE{\TYPE p \alpha {} ^D [\TYPE p {\hat\phi} {} ] =
(\alpha,\hat\phi)_\DB } When $\DB=\RR^n$ the bracket subscript will
be omitted.

If $  (\alpha,\hat\phi) $ is well defined \big(i.e. $\alpha$ is
integrable with respect to the bracket  $ ({.},{.}\big ) $ ) then one says that
$\alpha^D$ is a {\it regular } Schwartz $p-$form distribution
associated with $\alpha$. Not all distributions are regular and
associated with smooth forms.

Suppose $Z$ is a smooth vector field
on $\RR^n$ then $\widetilde Z \equiv  g(Z,-)$ is a smooth $1-$ form. Using the property
\BE{ \star\star =  t_g\, \eta^{n-1} } where the involution $\eta$ on forms satisfies $\eta\TYPE p \alpha {}
\equiv (-1)^p \TYPE p \alpha {}$ then it is straightforward to
verify the algebraic identities.
\BE{\star \TYPE q \alpha {} ^D [ \TYPE {n-q} \phi {} ]=
t_g\,\eta^{n-1} \TYPE q \alpha {} ^D [ \star^{-1} \TYPE {n-q} \phi
{} ]}
\BE{ ( \widetilde Z \wedge \alpha^D) [\phi]= \alpha^D[i_Z\phi] }
\BE{ i_Z \alpha^D[\beta]= \alpha^D[ \widetilde Z\wedge \beta] }
The operator $\delta\equiv\star ^{-1} \,d\, \star\eta$ is a formal
adjoint of the exterior derivative $d$ with respect to the bracket $({.},{.})$
in the sense that
 \BE{d\TYPE p \alpha{} [\TYPE {p+1} {\hat\phi}
{}]=\alpha[\delta\hat\phi]}
 Similarly
\BE{\delta \TYPE p\alpha {} [ \TYPE {p-1} {\hat\psi} {}] =
\alpha[d\hat\psi] }
These relations are used to define $d\,\alpha^D$ and $\delta\,\alpha^D$. If $f$ is a smooth $0-$form then we denote by $f\,\alpha^D$ the distribution defined by $f\,\alpha^D[\hat\chi]= \alpha^D[f\,\hat\chi]$.

If one defines the Lie derivative of $\TYPE p {\alpha^D} {}$ with respect to a smooth vector field $Z$ by
$${\cal L}_Z \,\TYPE p {\alpha^D} {} = (i_Z\,d + d\, i_Z)\,\TYPE p {\alpha^D} {}$$
then
\BAE{
{\cal L}_Z\,\TYPE p {\alpha^D} {} [\TYPE p {\hat\chi} {}]=\TYPE p {\alpha^D} {}[\delta\widetilde Z \wedge \TYPE p {\hat\chi} {}  + (\widetilde Z^\eta + \widetilde Z  )\,\delta \TYPE p {\hat\chi} {}  - \nabla_Z \TYPE p {\hat\chi} {}]
\label{LIE}}
where $\nabla$ denotes covariant differentiation with respect to the Levi-Civita connection for which $\delta\equiv - g^{ab}\,i_{X_a}\nabla_{X_b}$.

In terms of $d$ and $\delta$  one
defines \footnote{The traditional Laplacian operator on forms is  $-\Delta$}  the Hodge de Rham operator $\Delta= d\,\delta + \delta\,d$.
A number of important identities follow by integrating the Leibnitz
relation involving smooth forms
$$d(\alpha\wedge\beta)=d\alpha\wedge\beta+(\eta\alpha)\wedge
d\beta$$ over regular domains $\DB$. These include \BAE{{ \BRK
{d\phi} {\psi}   = \BRK{\phi}{\delta\psi} +
\int_{\partial\DB}\phi\,\wedge \star\,\psi}\label{key1}} \BAE{{ \BRK
{\delta\phi}  {\psi}   = \BRK{\phi}{d\psi} -
\int_{\partial\DB}\psi\,\wedge \star\,\phi}\label{key2}} \BAE{ \BRK
{\Delta\phi} {\psi} - \BRK {d\phi} {d\psi} -\BRK
{\delta\phi}{\delta\psi}= \int_{\partial\DB}(\delta\phi\,\wedge
\star\,\psi -\psi\,\wedge \star\, d\phi)\label{key3}} \BAE{ \BRK
{\Delta\phi} {\psi} - \BRK {\phi} {\Delta\psi}=
\int_{\partial\DB}\left(  \delta\phi\wedge \star\,\psi
-\delta\psi\wedge\star\,\phi +\phi\wedge \star\, d\psi -\psi\wedge \star\,
d\phi \right)\label{key4}}
 From these elementary identities further
useful identities involving singular double-forms can be obtained.

\section{Double-Forms on $\RR^n$}
\label{ch2}
The notion of a double-form was introduced in order to analyse problems in potential theory in ${\bf R}^3$ \cite{adcalc}. Double-forms have since been used by Duff  \cite{duff} and others as a powerful tool to study exterior equations on more general manifolds.
A double-form of bi-degree $(p,q)$  over the product manifold
$X\times Y$ may be regarded for each point in $X$  as a $p$-form
valued $q-$form on $Y$ or for each point in $Y$  as a $q$-form
valued $p-$form on $X$  .
To illustrate these notions suppose $X=Y=\RR^n$. In the natural
coordinate system $(x^i)$ for $X$ and $(y^i)$ for Y one has\footnote{If $f:{\bf R}^n \times {\bf R}^n \mapsto {\bf R}$ we denote its image by $f(X,Y)$ with a similar notation for double-forms}:
\BE{
\TYPE {p,q} {\gamma} {}  (X,Y)   = \sum_{i_1<\ldots <i_p}\sum_{j_1<\ldots <j_q} \gamma_{i_1\ldots i_p, j_1\ldots j_p}(X,Y) \epr  x i p \odot \epr  y j q
}
The {\it symmetric}  product $\odot$ is defined so that:
\BE{
 ( d\, x^i \odot  d\, y^j)  \wedge\, (  d\, x ^k \odot d\, y ^s  )= ( d\, x^i \odot  d\, x^k)  \wedge\, (  d\, y ^j \odot d\, y ^s  )
}
and this implies
\BE{
\TYPE {p,q} {\gamma} {} \wedge \,  \TYPE {\pp,\qq} {\gamma^\prime} {} = (-1)^{p\pp+q\qq}\,\TYPE {\pp,\qq} {\gamma^\prime} {}\wedge\,\TYPE {p,q} {\gamma} {}
}
for all $\gamma$, $\gamma^{\prime}$.
The Euclidean metric tensors in natural coordinates and their associated Hodge maps will be designated:
\BE{g_X=\sum_{i=1}^n d\,x^i \otimes d\,x^i }
\BE{g_Y=\sum_{i=1}^n d\,y^i \otimes d\,y^i }
and \BE{ \STAR X 1 = dx^1\wedge dx^2\wedge\ldots\wedge d x^n }
\BE{ \STAR Y 1 = dy^1\wedge dy^2\wedge\ldots\wedge d y^n }
The operators
$d_X,d_Y,\delta_X,\delta_Y,\STAR X,\STAR Y$ then act naturally on
forms in the appropriate space.

The {\it fundamental double-form of bi-degree} $(p,p)$ is defined as
\BE{\TYPE {p,p} {\gamma}  {} (X,Y) =\sum_{i_1<i_2\ldots <i_p}f(X,Y)
( d\,x^{i_1}\wedge \ldots \wedge d\,x^{i_p}  )_X \odot
(d\,y^{i_1}\wedge \ldots \wedge d\,y^{i_p})_Y} where
 \BE{
 f(X,Y)= \begin{cases}
 \frac{1}{(n-2)V_{n-1}}\frac{1}{\vert \xx -\yy
\vert^{n-2} } & n>2
\\ \qquad\\\frac{1}{2\pi}\log \frac{1}{\vert \xx -\yy \vert} & n=2
\end{cases}
 }
 and \footnote{$V_n$ is the volume of the unit $n$-sphere in ${\bf R}^n$.  }
 \BE{V_{n-1}=V_{n-2}\int_0^\pi
\sin^{n-2}\theta\,d\,\theta } with $V_1=2\pi,\,\,V_0=1$ and $ \vert \xx - \yy
\vert^2= \sum_{i=1}^n (x^i -y^i)^2$. This double-form is singular on
$X\times Y$ when $\xx=\yy$ and satisfies \BE{\gm p = \TYPE {p,p}
\gamma {} (Y,X)   } \BE{\Delta_X \gm p =0\qquad\qquad X\neq Y} \BE{
d_X  \gm p = \delta_Y \gm {p+1} } \BE{ \delta_X \gm {p+1} =d_Y \gm p
} \BE{ \Delta_X \gm p = \Delta_Y \gm p }

The singularity at  ${\bf x}={\bf y} $ in $\gm p$ permeates many of
the integrands of integrals that feature below. Such integrals, when they exist, are then defined
implicitly as limits. In general integrals $\int_\DB\alpha$ of
$n$-forms $\alpha$, over a domain $\DB\subset \RR^n$, that contain such a
singularity are understood as
$$\TYPE {\epsilon\to 0} {Lim} {} \,\,\int_{\DB -{B_\epsilon}} \alpha$$
where ${B_{\epsilon}}$ is an $n$-ball of radius $\epsilon$ centred on the
location of the singularity.

For any smooth $p-$form $\alpha$ on $Y$ the double-form $\gm p$ may
be used to define the type-preserving integral operator $\Gamma$:
 $\alpha_X\to (\Gamma\alpha)_Y$ mapping $p-$forms on $X$ to  $p-$forms on $Y$:
where \BE{ (\Gamma\alpha)_Y=\int_X \alpha_X \wedge \STAR X \gm p
=\int_X \gm p \wedge \STAR X \alpha_X}
 From (\ref{key1})  and  (\ref{key2})  it follows that  for all test forms $\hat\phi_X$, the differential operators
$d_X$ and $\delta_X$ commute with the integral operator $\Gamma$:
\BE{ d_X(\Gamma \hat\phi_X)=\Gamma(d_X\hat\phi_X)  } \BE{\delta_X
(\Gamma \hat\phi_X)= \Gamma(\delta_X\hat\phi_X)   }
If one writes
 \BE{
(\Gamma{\TYPE p \phi {}})_Y=\left(\gm p, \phi_X\right)} and notes
that $\Gamma$ is a symmetric operator:
$$(\Gamma\alpha,\beta)=(\alpha,\Gamma\beta)=(\Gamma\beta,\alpha)$$
then it follows from Green's identity applied to forms with compact
support that: \BE{ \hat\psi=\Gamma\Delta\hat\psi }
Thus \BE{(\Gamma\phi,\Delta\hp) = (\phi,\Gamma\delta\hp) =
(\phi,\hp) } But $ (\Gamma\phi,\Delta\hp)=(\Delta\Gamma\phi,\hp) $
hence \BE{ (\Delta\Gamma\phi,\hp)=(\phi,\hp) } or
\BE{\Delta\Gamma\phi[\hp]=\phi[\hp]} and $\Gamma$ is seen to be an
inverse of $\Delta$.

For boundary value problems it is useful to introduce singular
double-forms on $\DB$ that differ from the fundamental double-forms
$\gm p$ by regular solutions to Laplace's equation $\Delta\TYPE p
\phi {}=0$. Such double-forms inherit the singularity structure of
$\gm p$ but can  be tailored to satisfy certain conditions on the
boundary $\partial\DB$.

As an example suppose one requires $p-$form solutions to the generalized Poisson
equation
\BE{\Delta\TYPE p \phi {}= \TYPE p {\cal S} {}\label{poisson}}
 that vanish on $\partial\DB$. Denote by \BE{\GGG^p_{XY}=\gm
p -\widetilde{\gm p}} with $ \widetilde{\gm p} $ a symmetric {\it
non-singular solution} of $\Delta \widetilde{\gm p}=0$ that
coincides with the singular solution $\gm p$ on $\bDD$. Thus
\BE{\Delta\GGG^p_{XY}=0\qquad X\neq Y} \BE{\GGG^p_{XY}=\GGG^p_{XY}}
and \BE{\GGG^p_{XY}\vert_{\bDD}=0}

Suppose more generally  now that $\GGG^p_{XY}$ is a double-form
satisfying chosen conditions on $\bDD$. For any smooth $p-$form on
$\RR^n$ define the integral operator $\GGG$ by
\BAE{(\GGG\phi)_Y=\left( \GGG^p_{XY},\phi_X \right)\label{GGG} } If
${\bf y}\in \DB$ the integral is singular at ${\bf x}={\bf y}$ and
\BAE{\left( \GGG^p_{XY},\phi_X\right)=\TYPE {\epsilon\to 0} {Lim}
{}\,\, \left( \GGG^P_{XY},\phi\right)_{\DB-B_{\epsilon}}  }
One may use (\ref{key4}) together with the limits (derived by
explicit computation in spherical polar coordinates with origin at
the center of $B_\epsilon$):
$$\TYPE {\epsilon\to 0} {Lim} {} \,\,\int_{S_\epsilon} \left( \delta_X\GGG^p_{XY}\wedge \TYPE X \star {}\phi_X
 -\phi_X\wedge \TYPE X\star {} d_X\GGG^p_{XY}   \right)=\nu\phi_Y$$
$$\TYPE {\epsilon\to 0} {Lim} {} \,\,\int_{S_\epsilon}
\left( \delta_X\phi_X\wedge\TYPE X\star {}\GGG^p_{XY} -
\GGG^p_{XY}\wedge \TYPE X\star{} \,d_X\phi_X  \right)=0 $$ where
$S_\epsilon$ is the $n-1$ dimensional surface of the ball $B_\epsilon$, to verify
\BAE{&\nu\phi_Y=(\GGG\,\Delta\phi)_Y \\&-\int_\bDD\left(
(\delta\phi)_X\wedge       \TYPE X \star {}          \GGG^p_{XY}
-\GGG^p_{XY}\wedge    \TYPE X \star {}     (d\phi)_X
-\delta_X\GGG^p_{XY}\wedge   \TYPE X \star {}      \phi_X
+\phi_X\wedge   \TYPE X \star {}     \GGG^p_{XY} \right)\label{key5} }
In this expression, $\nu=0$ if $\bf y$ is outside $\DB$, $\nu=\frac{1}{2}$ if $\bf y \in \partial\DB$
and $\nu=1$ if $\bf y\in\DB$
If the first
term on the right of this equation is written in terms of the source
$\TYPE p {\cal S} {}$ in the equation (\ref{poisson})  then (\ref{key5})
offers a representation of the solution of this equation in terms of
this source and values of $\delta\phi$, $d\phi$, $\star\phi$ and $\phi$ on
$\bDD$. It should be stressed however that this representation is
not an explicit solution in general since
$\delta\phi$, $d\phi$, $\star\phi$ and $\phi$ cannot in general be assigned values arbitrarily on $\bDD$.
 Nevertheless this representation
is the cornerstone of many developments in potential theory \cite{jackson}.  The
above identities and their subsequent uses are often attributed to
Green.

\section{Electromagnetic Fields in Spacetime}
\label{ch_fields}

Maxwell's equations for an electromagnetic field in an arbitrary
medium can be written
\begin{align}
d\,F=0 \qquadand d\,\star\, G =j, \label{Maxwell}
\end{align}
where $F\in\GamLamM{2}$ is the Maxwell $2$-form, $G\in\GamLamM{2}$ is
the excitation $2$-form and $j\in\GamLamM{3}$ is the $3$-form electric
current source\footnote{All tensors in this article have dimensions
  constructed from the SI dimensions $[M], [L], [T], [Q]$ where $[Q]$
  has the unit of the Coulomb in the MKS system. We adopt $[g]=[L^2],
  [G]=[j]=[Q],\,[F]=[Q]/[\ee]$ where the permittivity of free space
  $\epsilon_0$ has the dimensions $ [ Q^2\,T^2 M^{-1}\,L^{-3}] $ and
  $c_0=\frac{1}{\sqrt{\ee\mu_0}}$ denotes the speed of light in vacuo.
  Note that, with $[g]=[L^2]$, for $p-$forms $\alpha$ in $n$ dimensions one has $[\star \alpha]=[\alpha] [L^{n-2p}]$}.
  To close this system, ``electromagnetic constitutive
relations'' relating $G$ and $j$ to $F$ are necessary.

The electric $4$-current $j$  describes both  (mobile) electric charge
and effective (Ohmic) currents in a conducting medium.  The {\it
electric } field $\Me\in\GamLamM{1}$ and {\it magnetic induction
field $\Mb\in\GamLamM{1}$} associated with $F$ are defined with
respect to an arbitrary {\it unit} future-pointing timelike
$4$-velocity vector field $U\in\GamTM$ by
\begin{align}
\Me = \iuF \qquadand \cc\Mb = \iustarF. \label{intro_e_b}
\end{align}
Thus $i_U\Me=0$ and $i_U\Mb=0$ and since $g(U,U)=-1$
\begin{equation}
F=\Me\wedge \dualu - \star\,(\cc\Mb\wedge \dualu). \label{intro_F}
\end{equation}

The field $U$ may be used to describe an {\it observer frame} on
spacetime and its integral curves model idealized  observers.

Likewise the {\it displacement} field $\Md\in\GamLamM{1}$ and the
{\it magnetic} field $\Mh\in\GamLamM{1}$  associated with $G$ are
defined with respect to $U$ by
\begin{align}
\Md = \iu G\,, \qquadand \Mh/\cc = \iu\star G\,. \label{Media_d_h}
\end{align}
Thus
\begin{equation}
G=\Md\wedge \dualu - \star\,((\Mh/\cc)\wedge \dualu), \label{Media_G}
\end{equation}
and $i_U\Md=0$ and $i_U\Mh=0$.
\section{Time dependent Maxwell Systems in Space}
\label{ch time}

In this article we restrict to fields on Minkowski spacetime which can be globally foliated by 3-dimensional spacelike hyperplanes. The Minkowski metric induces a  metric  with Euclidean signature on each hyperplane. Furthermore each hyperplane contains events that are deemed simultaneous with respect to a clock attached to any integral curve of a future-pointing unit time-like vector field  $U=\frac{1}{c}\frac{\partial}{\partial t}$ on spacetime and the Hodge map $\star$ induces a Hodge map $\HASH$ on each hyperplane by the relation $$\star 1= c\, d\, t \wedge \HASH 1$$
The spacetime Maxwell system can now be reduced to a family of exterior systems on  ${\bf R}^3$. Each member is an exterior system involving forms on ${\bf R}^3$ depending parametrically on time $t$.
 Let the $3+1$ split  of the 4-current $3$-form with respect to the foliation
be
\BE{
{\TYPE 3 j {}}= - {\TYPE 2 \bbfJ {}} \wedge d\,t + \TYPE
0 \rho {}  \hash 1,
}
with $i_{\PD t} {\TYPE 2 \bbfJ {}}=0 $. Then, from
(\ref{Maxwell})
\BE{
d\, j=0,\label{dj}
}
yields
\BE{
\D {\TYPE 2 \bbfJ {}} + \dot{\underset{(0)}\rho} \hash 1=0.\label{cont}
}
Here and below an over-dot denotes (Lie) differentiation with respect to the parameter $t$ and $\dot\alpha\equiv {\cal L}_{\frac{\partial}{\partial t}}\alpha$.
It is convenient to introduce on each hyperplane the (Hodge) dual forms:
$$
\bfE:=\hash \bfe, \quad \bfD:=\hash\bfd,\quad \bfB:=\hash \bfb, \quad
\bfH:=\hash \bfh,\quad \TYPE 1 {\bbfj} {}:=\hash\TYPE 2 {\bbfJ} {}
$$
so that the $3+1$ split of the spacetime covariant Maxwell
equations (\ref{Maxwell})  with respect to $\widetilde U=-c\,d t$ becomes
\BE{\D\bfe=-\bfBdot,\label{M1}}
\BE{\D\bfB=0,\label{M2}}
\BE{\D\bfh={\TYPE 2 \bbfJ {}}+\bfDdot,\label{M3}}
\BE{\D\bfD=\underset{(0)}\rho\hash 1.\label{M4}}

All $p$-forms ($p\ge 0$) in these equations are independent of $dt$
but have components that may depend parametrically on $t$.

In the following it is assumed that  $\bfb=\mu \bfh$ and $ \bfd=\ep
\bfe$   where $\ep=\EP$, $\mu=\MU$. Thus
in terms of $\bfe,\bfh, \bfE, \bfH$:
\BE{\D   \bfe=-\mu\bfHdot,\label{M11}}
\BE{\D   \bfH=0,\label{M12}}
\BE{\D   \bfh=\ep \bfEdot + {\TYPE 2 \bbfJ {}},   \label{M13}}
\BE{\ep\D\bfE=\underset{(0)}\rho \hash 1.\label{M14}}
It is straightforward to show from this exterior system that the fields $\bfe$ and $\bfh$ satisfy
\BAE{
\DDelta\,\bfh + \epsilon\mu \ddot\bfh=\Ddelta\bbfJ + \hash\left(\D\,\epsilon\wedge \hash\epsilon^{-1}\left( \bbfJ - \D\,\bfh   \right)       \right)\label{ewave}
}
\BAE{
\DDelta\,\bfe + \epsilon\mu \ddot\bfe=-\D\left(\frac{\rho}{\epsilon}\right)-\dot\bbfj +\hash\left(\D\,\mu\wedge
 \hash\left( \mu^{-1}\,\D\,\bfe \right)       \right)   \label{hwave}
}
Thus for homogeneous  media with zero conductivity these equations each reduce to driven wave equations. Henceforth  $\ep, \mu $ are assumed to be constant scalars.

Locally on spacetime one has an equivalence class of $1-$forms whose elements $A$ differ by the addition of any exact $1-$form. Then since $F$ is closed one may write locally  $F=d\,A$. Decomposing $A$ (with respect to a particular $\widetilde U$)  into $0-$form $\phi$ and 1-form $\bfA$  potentials yields
\BE{A=-\phi\,\widetilde U - c\bfA
}
with $i_U\bfA=0$. With $\widetilde U=-c\,d\,t$ and $\star 1= \HASH 1 \wedge \tilde U   $  one has in terms of $t-$parameterised forms on ${\bf R^3}$
$$\bfe=-\D\phi-\dot\bfA$$
$$\bfb=\hash\,\D \bfA$$
The Maxwell system then reduces to
\BAE{
\Ddelta\,\D\phi + \Ddelta \dot\bfA= \frac{\rho}{\epsilon}
}
\BAE{
\Ddelta\, \D\bfA + \epsilon\mu \ddot \bfA = \mu \,\bbfj - \epsilon\mu \D\,\dot\phi
}

In a gauge with $$\Ddelta\bfA-\epsilon\mu\dot\phi=0$$ 
the Maxwell system above requires that the potentials must satisfy
\BE{
\DDelta \bfA+\epsilon\mu\,\ddot\bfA=\mu\,  \bbfj
\label{MAX1}}
\BE{
\DDelta \phi+\epsilon\mu\,\ddot\phi=\frac{\rho}{\epsilon}
\label{MAX2}}
for sources satisfying (\ref{cont}). Physical solutions correspond to those satisfying physically motivated boundary conditions (in both space and time).
Alternatively in a gauge with \BE{\Ddelta\bfA =0\label{coulomb}} the Maxwell system becomes
\BAE{
\DDelta\phi=\frac{\rho}{\epsilon}
}
\BAE{
\DDelta\bfA+\epsilon\mu \ddot\bfA=\mu\bbfj - \epsilon\mu\,\D\,\dot\phi
}
In these equations on ${\bf R^3}$ with the Euclidean metric, the Hodge map on all forms satisfies $\HASH\HASH=1$ so $ \Ddelta=\hash \D\hash\,\eta $ and $\DDelta=\hash\D\hash\,\eta\D + \D \,\hash\D\hash\eta$. Henceforth,  attention is devoted mainly to exterior systems on ${\bf R^3}$ so it is unnecessary to maintain the notational distinction between $\D$ and $d$, $\Ddelta$ and $\delta $, $\DDelta$ and $\Delta$.

\section{Electrostatics in $\bf R^3$  with Sources in Domains with Boundaries}
\label{ch3}

\def\Tilde{\bar}

\def\rhoo{\bar\rho}

\newcommand\TTilde[1]{{#1}^\prime}


Integral operators analogous to  $\Gamma$ may be used to solve electrostatic
boundary value problems in ${\bf R^3}$ involving the Hodge-de Rham operator $\Delta$ above and electric fields independent of $t$. As a basic
example consider the Dirichlet  problem of finding the electrostatic
$0-$form potential $\phi$ on $\DD\subset\RR^3$ that must satisfy
 \BE{
\begin{cases}
\Delta\phi={\TYPE 0 {\bar\rho} {}} & {}\\
\phi\vert_{\partial\DD}=\Tilde\phi
\end{cases}
} for some data $\rhoo\equiv \frac{\rho}{\epsilon} \in \DD,\Tilde\phi\in \bDD$.

 The solution may be represented  in terms of the
integral operator $\GGG$ defined by
 \BE{ (\GGG\rhoo)_Y= \int_{\DD(X)}
  \rhoo_X \HASHX  \GGG_{XY}
} and is given by  \BE{
 \phi_Y=(\GGG\rhoo)_Y
  - \int_\bDD \Tilde\phi_X
\HASHX d_X \GGG_{XY}
 }

where \BE{\GGG_{XY}=\gm 0 -\widetilde{\gm 0}} with $ \widetilde{\gm
0} $ a symmetric {\it non-singular solution} of $\Delta
\widetilde{\gm 0}=0$ that coincides with the singular solution $\gm
0$ on $\bDD$. Thus \BE{\Delta\GGG_{XY}=0\qquad X\neq Y}
\BE{\GGG_{XY}=\GGG_{YX}} and \BE{\GGG_{XY}\vert_{\bDD}=0}

If the data is smooth it defines {\it regular} distributions
$\rhoo^D, \Tilde{\phi}^D $ with supports on $\DD$ and $\bDD$
respectively by \BE{ \rhoo^D[\hp]=(\rhoo,\hp)\qquad \qquad
\Tilde{\phi}^D[\hp]=-\int_{\bDD(X)} \Tilde{\phi}_X \HASHX d_X(\GGG
\hp)_X } Now \BE{
(\GGG\rhoo)[\hp]=(\GGG\rhoo,\hp)=(\rhoo,\GGG\hp)=\rhoo[\GGG\hp] }
 Thus for  {\it singular} distributional sources $\rhoo^D$ one then
has distributional Dirichlet solutions: \BE{
\phi^D[\hp]=\rhoo^D[\GGG\hp]+ \Tilde{\phi}^D[\hp] } If the
eigenvalues $\lambda_M$  and Dirichlet eigenfunctions $\Phi_N$ of
$\Delta$ on $\DD$ can be found by solving \BE{
\begin{cases}
\Delta\Phi_M=\lambda_M \Phi_M&\\\qquad\\ \Phi_M\vert_\bDD=0
\end{cases}
} then a traditional way to satisfy the conditions required of
$\GGG_{XY}$ is to express it as a Fourier expansion in  the
eigenmodes $\Phi_M(Y)$ for each point of $Y\in \DD$.

If the real modes are labeled by a discrete index set $M$ the
expansion takes the form
 \BE{\GGG_{XY}= \sum_M \GGG_M(X) \Phi_M(Y) }
If such modes (with support on $\DD$) are ortho-normalised so that
$({\Phi_M},\Phi_N)=\delta_{MN}$ then
\BE{\GGG_N(X)=(\GGG_{XY},\Phi_N(Y))}

Since the solutions $\Phi_N$ are zero on $\bDD$ it follows
that \BE{\Phi_N(Y)=\lambda_N\int_\DD \Phi_N(X) \HASHX
\GGG_{XY}=\lambda_N(\Phi_N(X), \GGG_{XY}) =\lambda_N\GGG_N(Y)
 }
or $$\GGG_N(X) = \frac{1}{\lambda_N}
 \Phi_N(X)\qquad \lambda_N\neq
0$$

Thus \BE{  \GGG_{XY}=\sum_N  \frac{1}{\lambda_N}
\Phi_N(X)\,\Phi_N(Y) } This series must be regarded as weakly
convergent and the summations become integrations when the
eigenvalues are continuous.

For example consider $\DD$ to be the space between two perfectly
conducting parallel plates at $z=0$ and $z=L$, separated by a distance $L$ in vacuo.
Taking cartesian coordinates $\{x,y,z\}$ for $X$ and
$\{\xp,\yp,\zp\}$ for $Y$ the eigenvalues $N=\{k_x,k_y,n\}$ with
$-\infty<k_x<\infty$, $-\infty<k_y<\infty$, $n=1,2,\ldots$ and (in
complexified form): \BE{ \Phi_N(x,y,z)=
\frac{1}{2\pi}\sqrt{\frac{2}{L}} \exp(ik_x +ik_y)\, \sin\frac{n\pi
z}{L} }
 This gives
\BAE{ \GGG(x,y,z;\,\xp,&\yp,\zp)= \\&\frac{2}{4\pi^2 L}
\sum_n^\infty\sin\frac{n\pi z}{L}  \sin\frac{n\pi
\zp}{L}\int_{-\infty}^\infty\int_{-\infty}^\infty\frac{dk_x\,dk_y\,
e^{ik_x(x-\xp) +ik_y(y-\yp)}}{ k_x^2 + k_y^2 + \frac{n^2\pi^2}{L^2}
}\label{GSUM} } If the source $\rhoo$ is not smoothly distributed in $\DD$ one
needs further technology to evaluate the solution  and this will be
developed below.

\section{Magnetostatics in $\RR^3$ with  Smooth Sources}
\label{ch4}

Since magnetic charge is absent in Nature a typical problem in
magnetostatics is the determination of a static magnetic field from
a stationary electric current. The magnetostatic equations are a
subset of Maxwell's equations and may be written: \BE{\delta
\bfh  =0\label{mag1}} \BE{d\,\bfh=\TYPE 2 {\bf J} {}\label{mag2} } in terms of forms on $\RR^3$. If the
current is localized in space (vanishing at $\infty$) the boundary
conditions can be accommodated by using $\gm 1$ for the solution.
If one sets \footnote{Since $d \TYPE 2 {\bf B}{} =0$ one has $\TYPE 2 {\bf B}{} =d\TYPE 1 {\bfA} {}$ in
regular domains and in vacuo $\hash {\bf B}=\mu_0 {\bf h}$ where $\mu_0$ is the
permeability of free space.} ${\bf h}=\hash d \bfA/\mu_0$ then   (\ref{mag1}) is immediately satisfied.
Writing  $\TYPE 1 {\JJJ} {}= -\mu_0\hash {\bf J}$  one must have from (\ref{mag2}) \BE{d\HASH
\JJJ=0} and in the gauge with $\delta \bfA=0$  equation (\ref{mag2}) becomes 
\BE{\Delta\bfA= \JJJ} Thus in free space with a smooth source $\JJJ$ \BE{ \bfA = \Gamma\JJJ } since
\BE{ \Delta\bfA=\Delta\Gamma\JJJ=\JJJ } i.e. \BE{ \bfA_Y=\int_X\,\gm
{1}  \wedge \HASHX \JJJ  } with $$ \gm 1 = \frac{1}{4\pi\,\vert \xx - \yy
\vert} \sum_{j=1}^3 \, dx^{j}\odot \,d{x^\prime}^j$$

Note if $\JJJ$ is bounded
 \BAE{
\delta_Y\bfA_Y=\int_X \delta_Y \gm 1 \wedge \HASHX\JJJ_X=\int_X
d_X\gm 0\wedge \HASHX \JJJ_X\\ =-\int_X \gm 0\, d_X \HASHX\JJJ_X =0}
 Since
$\Delta \TYPE 1 {\bf h} {} = \delta {\bf J}$ one also has directly
\BE{\TYPE 1 {\poph} {}(Y)= \int_X \gm 1
\wedge \HASHX (\delta \bbfJ)_X.\label{aaa} }
That this also furnishes a solution to (\ref{mag1}) and (\ref{mag2}) will be verified in section [\ref{ch6}].

Now
 for a  smooth (regular) source $\JJJ$
\BAE{ \bfA[\TYPE  1{\hat\psi}
{}]=\Gamma\JJJ[{\hat\psi}]=\int_Y(\Gamma\JJJ)_Y\wedge
\HASHY{\hat\psi}_Y\\ =\int_Y\left( \int_X \gm 1 \wedge \HASHX \JJJ_X   \right) \wedge \HASHY {\hat\psi}_Y\\
=\int_X\int_Y \left( \gm 1 \wedge \HASHY {\hat\psi}_Y\right) \wedge
\HASHX
\JJJ_X\\
=\int_X(\Gamma{\hat\psi})_X \wedge \HASHX \JJJ_X\\=\int_X\JJJ_X
\wedge \HASHX (\Gamma{\hat\psi})_X\\\equiv\JJJ^D[\Gamma {\hat\psi}]\label{step}}

If $\JJJ$ is a smooth $1-$form in $\RR^3$ the integral
$\JJJ^D[\Gamma {\hat\psi}]  $ furnishes a solution for $\bfA$.
However if the current has support on a curve in $\RR^3$ it must be
regarded as a {\it distributional $1-$ form source} not associated
with a smooth $1-$form. Similarly the charge density in the
electrostatic problem may be restricted to a curve or surface in
$\RR^3$ in which case a distributional source $\rhoo^D$ not
associated with a smooth $0-$form must be specified.

\section{Dirac Distributions on Submanifolds}
\label{ch5}

\newcommand\ff{{f^{(r)}}}

To define singular distributions with support on (possibly disjoint) submanifolds  of a manifold $M$ it is convenient to describe each submanifold parametrically as an embedding in $M$. Suppose that a distribution has support on  a collection of submanifolds $S_0^{(r)}$,  i.e on the chain $ \sum_r S_0^{(r)}$.  Recall that an
$n-k$ dimensional submanifold  $S_0^{(r)}$ (possibly with boundary)  in an $n-$dimensional manifold
$M$ can also be prescribed in terms of $k$ $0-$forms
$\ff_1,\ff_2,\ldots,\ff_k$ on $M$, such that $  d\ff_1\wedge
d\ff_2\ldots\wedge d\ff_k\neq 0$. Such forms generate a local foliation
$S^{(r)}$  in the neighborhood of each submanifold $S_0^{(r)}$, where each leaf  $S_{\bf c}^{(r)}  $ of the foliation is an $n-k$ dimensional
embedding given by $\ff_1=c_1,\ff_2=c_2,\ldots,\ff_k=c_k$ for some
constants $c_1,c_2,\ldots c_k$ and ${\bf c}=(c_1,c_2, \ldots c_k)$.
 We may choose the leaf $S_0^{(r)}$  with all
these constants zero and denote it $ \Sigma_{\bf \ff}^{n-k} $ with ${\bf
\ff}\equiv ( \ff_1,\ff_2,\ldots,\ff_k) $. It is sufficient to establish the distributional framework for a single component of a general chain so henceforth the label $(r)$ will be omitted.

If $M$ is endowed with a metric $g$ (and associated Hodge map $\star$), the forms $\{df_j\}$ give rise to a class of forms $\TYPE {n-k}
{\Omega_{\mathbf f}} {}$ on $M$ defined with respect to $\star 1$ by
 \BE{\star 1 = df_1\wedge df_2\wedge\ldots \wedge df_k \wedge \TYPE
{n-k} {\Omega_{\mathbf f}} {}\label{om}}

This class restricts to a natural class of measures on $\SIG {n-k}$.
 Thus each representative induces the
measure \BE{\TYPE {n-k} {\omega_{\mathbf f}} {}=  \TYPE {n-k}
{\Omega_{\mathbf f}} {}\vert_{\Sigma_{\mathbf f}^{n-k}} } on the
submanifold  $\SIG {n-k}$. Members of the class on $M$ are
equivalent if they differ by any combination of the $df_j$. This
{\it gauge freedom} is of no significance on $\SIG {n-k}$ since the
$df_j$ vanish there under pull-back. Thus it is sufficient to
represent the class by an element satisfying \BE{ i_{df_j} \TYPE
{n-k} {\Omega_{\mathbf f}} {}=0 \qquad\qquad j=1\ldots k .}  In
those situations where the leaves form orthogonal families, i.e.  $
g^{-1}(d f_i,d f_j)=0$ for $i\ne j$ then
 \BE{\BIGLERAY {n-k}=
\frac{\star\, (df_1\wedge df_2\wedge \ldots \wedge
df_k)}{\Pi_{j=1}^k \vert df_j\vert^2} } where $\vert
df_j\vert^2=\widetilde{df_j}(df_j)\neq 0$ and the sign of $\vert df_j\vert$ is
chosen so that the direction of the vector field $\widetilde{df_j}\equiv
g^{-1}(df_j,-)$ is in the direction of increasing values of $f_j$.
Since $df_j\neq 0$ one has on $\SIG {n-k}$ two fields of \lq\lq unit
normals": $n_j=\pm \frac{df_j}{\vert df_j\vert}$.  Then with
$n_j=\frac{df_j}{\vert df_j\vert}$ \BE{\BIGLERAY {n-k}
=\star\,\Pi_{j=1}^k\left( \frac{n_j}{\vert df_j\vert}\wedge\right) }
For non-orthogonal leaves an $\Omega_{\mathbf f}$ can be  chosen
directly from (\ref{om}). This condition  is equivalent to the
equation
 \BE{ \pm 1 = \star (df_1\wedge df_2\wedge\ldots \wedge df_k \wedge \TYPE
{n-k} {\Omega_{\mathbf f}} {}) \label{om1}} which yields freedom to
choose $\TYPE {n-k} {\omega_{\mathbf f}} {}$ with non-singular
coordinate components when restricted to $\SIG {n-k}$.

The singular Dirac $0-$form with support on $\SIG {n-k}$ is denoted
$\DSIG {n-k}$
and defined by
\BE{
\DSIG {n-k}[\TYPE 0{\hat\phi} {}]
=
\int_{\SIG {n-k}}\, {\hat\phi}\,   \LERAY {n-k}
 }
Furthermore if $Y$ is a smooth vector field on $M$ one may define
the directional derivative  $\DIRACY{n-k}$ of $\DSIG {n-k}$ by:
\BE{ {\DIRACY {n-k}}\equiv Y\,\DSIG {n-k} = i_Y\,d\DSIG{n-k}  }
It follows that
 \BE{ {\DIRACY {n-k}}[\TYPE 0
{\hat\phi} {}]=-\int_{\SIG {n-k}}\LERAYY {n-k}
  (\hat\phi) }
where for orthogonal leaves:
 \BE{ \LERAYY {n-k}(\hat\phi) = \,\frac{1}{\Pi_{j=1}^k \,\vert
df_j\vert ^2} \,\,\,i_{\widetilde{d f_k}}\ldots  i_{\widetilde{d f_1}}\,
d(\star\, \hat\phi \widetilde Y) \vert_{\SIG {n-k}} }
Since $$   \delta( {\TYPE 0 f {} } (  \widetilde Y {\TYPE 0 \psi {} }))
= f \delta(\widetilde Y \psi) -\psi (Yf)
    $$
    \BAE{ \LERAYY {n-k}(\hat\phi) &= -(\hat\phi \,\delta \widetilde Y - Y\hat\phi)
    \star\,\left( \frac{df_1}{\vert df_1  \vert^2}\wedge  \ldots \wedge \frac{df_k}{\vert df_k  \vert^2}
    \right)\\ &=   -(\hat\phi \,\delta \widetilde Y - Y\hat\phi) \BIGLERAY
    {n-k} \vert_{{\SIG {n-k}} }
      }
Additional directional derivatives yield the distributions
$${\DIRACYY {n-k}}\equiv
Y^1Y^2\ldots Y^q \DSIG{n-k} $$
 For example
 \BE{\left( Y^1\,Y^2 \DIRAC {n-k}\right) [\TYPE 0 {\hat\phi} {}]=
\DSIG {n-k}
 [\delta
(\widetilde{Y_1} \wedge \delta(  \hat\phi {\widetilde{Y_2}})    ) ]
 }

If one regards any point $p_0 \in M$ as  a $0-$dimensional space
then the original singular Dirac distribution  $\TYPE 0
{\delta^D_{p_0}} {} $ with support on $p_0$ is defined by: \BE{
\TYPE 0 {\delta^D_{p_0}} {}[{\TYPE 0 {\hat\phi} {} }]=\hat\phi(p_0)
} It follows that \BE{ \DSIG 0= \LERAY 0 \,\TYPE 0 {\delta^D_{p_0}}
{} }

The above singular $0-$form Dirac distributions can be used to
construct singular $p-$form Dirac distributions with support on
submanifolds in various ways. Thus in a spherical polar chart for
$\RR^3$ a $1-$form {\it $0$-layer} Dirac distribution on the unit
$2-$sphere centred at the origin may be represented as
 \BE{ \DSIG 2
\,\left(\mu_\theta(\theta,\phi)\,d\theta  +
\mu_\phi(\theta,\phi)\,d\phi\right) } where $\mu_\theta$ and
$\mu_\phi$ are smooth functions on the sphere. A $1-$form {\it
$1-$layer} Dirac distribution on the sphere (associated with some
vector field $Y$ on $\RR^3$) may take the form
 \BE{ \DIRACY 2
\,\left(p_\theta(\theta,\phi)\,d\theta  +
p_\phi(\theta,\phi)\,d\phi\right) } for smooth $p_\theta$ and
$p_\phi$.  Similarly a $2-$form {\it $0-$layer} Dirac distribution on
the sphere takes the form \BE{ \DSIG 2 \,\,q(\theta,\phi)\,d\theta
\wedge d\phi  } and a $2-$form {\it $1-$layer} Dirac distribution
takes the form \BE{ \DIRACY 2 s(\theta,\phi)\,d\theta \wedge d\phi }
for smooth $q$ and $s$ respectively. More generally in  coordinates
$$(\underbrace{\xi^1,\ldots,\xi^k}_{n-p},
\underbrace{\sigma_1,\ldots, \sigma_p}_p )$$ adapted to a
$p-$dimensional submanifold in $\RR^n$, one has for  $j=0,1,\ldots
p$, the  $j-$form,   \lq\lq r-layer\rq\rq\,  singular Dirac
distribution  with support on  $\SIG {n-k}$ :


\BE{{\TYPE j {\cal J} {}} =   {\TYPE 0 \lambda {}}
 (\sigma^{1}\ldots
\sigma^{p} ) \,
\TYPE 0 {{{\delta} {}}^{Y_1\ldots Y_r}_{\SIG {n-k}}} {}
\left( \sum_{i_1<i_2<\ldots <i_j} q_{i_1\ldots i_j}(\sigma^{1}\ldots
\sigma^{p} )\,d\sigma^{i_1}\wedge\sigma^{i_2}\wedge \ldots \wedge
d\sigma^{i_j} \right) }







For a point electrostatic dipole with dipole moment $Z_X=p^i\frac{\partial}{\partial x^i}$
located at ${\bf x_0}\in X$ the electrostatic potential distribution $\phi_Y^D$ is given by the singular distributional source $\rho^D_Y$ where
\BAE{
\phi_Y^D[\hat\chi]=\rho^D_Y[\hat\chi]=\left({\cal L}_{Z_X}\,\TYPE {0,0} {\gamma}{} (X,Y)\right)  \,\delta^D_{\bf x_0}[\hat\chi]={\bf p} \cdot \nabla\left( \frac{1}{\vert {\bf y} - {\bf {x_0}}  \vert}\right) \hat\chi({\bf {x_0}})
}
\section{  Line and Surface Currents in $\RR^3$}
\label{ch5.5}

The source of a smooth magnetostatic field is the smooth current
2-form $J$  in $\RR^3$ and the total electric current passing
through a surface $S\subset {\RR}^3$ is $I[S]=\int_S J  $,  measured
in amps in MKS units. In some circumstances this current density
  may be concentrated in the vicinity of 1-dimensional
submanifolds $\Sigma_1\subset\RR^3$  or 2-dimensional submanifolds
$\Sigma_2\subset\RR^3$.  Electric current filaments on material
curves in space (wires) and current sheets in material surfaces in
space are idealizations of such localized sources and may be
modeled by singular  distribution-valued 1-forms with support on curves and
surfaces respectively. In these situations new physical current
densities are introduced so that the total current in a segment of
wire or a region of a surface is finite and measurable. These singular
distributional sources can be mathematically modeled in
terms of the distributions $ \delta_{{\Sigma^{r}_{\bff}}} $ for
$r=1,2$ and a vector field $W$ in $\RR^3$ with support on
${\Sigma^{r}_{\bff}}$.

For a wire described by the curve with image  $ {\Sigma^{1}_{\bff}}$
introduce the distributional 1-form \BE{\TYPE 1 {{\cal
I}^D_{{\Sigma^{1}_{\bff}}}} {}=I_0\,\, \TYPE 1 {\widetilde W} {}\,\,
\TYPE 0 {\delta_{{\Sigma^{1}_{\bff}}}} {} }
 where $I_0$ is a smooth
function on the wire. Hence
$$
{\cal I}^D_{{\Sigma^{1}_{\bff}}}\,[\TYPE 1 {\hat\psi} {}] =
\int_{\RR^3} \TYPE 1 {\hat\psi} {}\wedge J
$$
 and with ${\cal J}=\HASH J$:
\BAE{ {\cal I}^D_{{\Sigma^{1}_{\bff}}}\,[ {\hat\psi}]& =
\delta_{{\Sigma^{1}_{\bff}}}[ I_0\, i_W\hat\psi  ]=
\int_{\Sigma^{1}_{\bff}}\,I_0\,( i_W\hat\psi)\,\LERAY
1=\int_{\RR^3}\hat\psi\wedge J =\int_{\RR^3}\hat\psi\wedge\HASH
{\cal J} = {\cal J}^D[\hat\psi] \\ &=\int_{\Sigma^{1}_{\bff}}
\hat\psi\, I_0 \,(i_W\LERAY 1) \equiv
\int_{\Sigma^{1}_{\bff}}\hat\psi \TYPE 0 {\breve I} {}}
since $\hat\psi \wedge \LERAY 1$ is zero on $  {\Sigma^{1}_{\bff}}$.
 It does not make sense to ascribe
a physical dimension to $ {\cal I}^D_{{\Sigma^{1}_{\bff}}}  $ since
it is a functional not a value. However if $\DIM {{\cal
J}^D[\hat\psi]}$ denotes the physical dimension of $ {{\cal
J}^D[\hat\psi]} $ one has in MKS units
$$
\DIM {{\cal J}^D[-]}= \DIMM {J}{-} = \mbox{amp}\,\DIM {-}
$$
and $\DIM {{\cal J}}=\DIM {\HASH J}= \,${amp}\, m${}^{-1}$ with
$\DIM {I_0}\DIM {\widetilde W} \DIM {\LERAY 1}=\DIM J=\, $ amp.
The (smooth) 0-form density

 \BAE{\TYPE 0 {\breve I} {}(s)\equiv  I_0 (i_W \LERAY
1)\,\label{lambda}} belonging to $ \Gamma\Lambda^0\SIG 1 $
 is the total current (amps) flowing along
the wire  at any point $s\in\SIG 1$

In a similar way, for  the 1-form distribution on $\RR^3$: \BAE{
\TYPE 1 {{\cal K}^D_{  {{\Sigma^{2}_{\bff}}}        }}  {}
=\kappa_0\, \widetilde W\, \TYPE 0 {    \delta _{
{{\Sigma^{2}_{\bff}}} } }  {} } one has \BAE{ {\cal K}^D_{
{{\Sigma^{2}_{\bff}}} }\,[ \hat\psi ]&= \int_{\RR^3}\,\hat\psi\wedge
J  =\, \TYPE 0 {    \delta _{ {{\Sigma^{2}_{\bff}}} } }  {} [
\kappa_0 i_W \hat\psi ]\\&= \int_{\SIG 2}\,\kappa_0 (i_W\hat\psi)
\LERAY 2 = \int_{\SIG 2}\hat\psi \wedge i_W( \kappa_0 \LERAY 2
)=\int_{\SIG 2} \hat\psi\wedge\TYPE 1 {\breve\kappa} {}
  }
since $ \kappa_0 (i_W\hat\psi)\, \LERAY 2= \kappa_0 \hat\psi \wedge
(i_W\LERAY 2)$ on $\SIG 2$. Here $$\TYPE 1 {\breve\kappa} {} =\kappa_0\,
i_W(\LERAY 2) \in\Gamma\Lambda^1\SIG 2$$ is a smooth 1-form (measured in
amps) on the surface $\SIG 2$. If $C_1\in\RR^3$ is any space-curve
lying on this surface then $\int_{C_1}\TYPE 1 {\breve\kappa} {}$ is the
total current in amps crossing this curve in the direction $W$, on
$\SIG 2$. This may be compared with the definition above of the
total current (in amps), $\int_S J$, crossing the surface $S\in
\RR^3$ in the direction $\widetilde{\hash J}$. If
$C_1^{\star}\frac{\partial}{\partial s}$ is a {\it unit} tangent
vector  \footnote{i.e. in terms of the Euclidean metric tensor $g$
in $\RR^3$, $g( C_1^{\star}\frac{\partial}{\partial
s},C_1^{\star}\frac{\partial}{\partial s} )=1$} to $C_1$ then
$i_{\frac{\partial}{\partial s}}\TYPE 1 {\breve\kappa} {}$ is the {\it
surface current density} in amp $m^{-1}$.

It is worth noting that $I_0\in \Gamma\Lambda^0\RR^3,\,\kappa_0\in
\Gamma\Lambda^0\RR^3,\, W\in \Gamma T\RR^3$ and $\SIG 1,\,\SIG 2$ are the primary
objects used to define the distributions $ {\cal I}^D_{\SIG 1} $ and
$ {\cal K}^D_{\SIG 2}  $ in $\RR^3$ in terms of which are defined
the smooth 0-forms $\breve I\in\Gamma\Lambda^0\SIG 1$ and 1-forms
$\breve\kappa\in\Gamma\Lambda^1{\SIG 2}$.

\section{ Magnetostatic Fields from a Singular Distributional Stationary Current Source}
\label{ch6}

Suppose a uniform electric current confined to a wire  in $\RR^3$ is
modeled in terms of  a $1-$form $0-${\it layer} Dirac distribution
$\delta_{\SIG 1}$ with support on the wire. A distributional source
$\JJJ^D$ will generate a distributional $1-$form $\bfA^D$.
Thus define $\bfA^D$ by \BE{\bfA^D[{\hat\psi}]= \JJJ^D[\Gamma
{\hat\psi}]} Suppose $\JJJ^D$ has a support on the locus  described
by $\bff=(f_1,f_2)$ with  $f_1= r-a$, and $f_2=z$ in cylindrical
coordinates $(r,\phi,z)$. This will be used to model a uniform stationary current
source in a circular loop of radius $a$  if \BE{ \JJJ^D=\lambda
\DIRAC 1 \,d\phi} with some constant $\lambda$.
Thus \BAE{ \JJJ^D[{\Gamma\hat\psi}]=\lambda \DIRAC 1 [i_{\widetilde{d\phi}} (\Gamma {\hat\psi}) ]\\
=\lambda \int _{\SIG 1 }\,\, \left(i_{\widetilde{d\phi}} (\Gamma
{\hat\psi})_Y\right) \,\LERAYY 1 } Since \BE{
(\Gamma{\hat\psi})_Y=\int_X \gm 1 \wedge \HASHX {\hat\psi}_X } and
\BE{ i_{\widetilde{d\phi}(Y)}(\Gamma{\hat\psi})_Y=\int_X
\left(i_{\widetilde{d\phi}(Y)}\gm 1\right) \wedge \HASHX {\hat\psi}_X }
then \BE{\bfA^D[{\hat\psi}] = \int_X {\TYPE 1 {{\cal A}_X} {}}\wedge
\HASHX {\hat\psi}_X} where \BE{{\cal A}_X =\lambda \int_{\SIG
1(Y)}\left( i_{\widetilde{d\phi}(Y)}\gm 1\right)\,\, \LERAYY 1 }

In cylindrical polar coordinates,
$\xx=(r\cos\phi,r\sin\phi,z)$,
$\yy=(\rp\cos\phip,\rp\sin\phip,\zp)$
$$\vert\xx-\yy\vert^2\equiv R^2(r,\phi,z;\rp,\phip,\zp)=r^2+\rp^2 +z^2+\zp^2 -2z\zp -
2r\rp\cos(\phi-\phip)$$
and the Euclidean metric
tensor is \BAE{g=d\, r \otimes d\,r + r^2\,d\,\phi \otimes d\,\phi +
d\,z\otimes d\, z} with
 \BAE{g^{-1}=\pd r
\otimes \pd r + r^{-2}\pd \phi \otimes \pd\phi + \pd z\otimes \pd z}

Hence ${\widetilde{d\phi^\prime}}=\frac{1}{{r^\prime}^2}\pd
{\phi^\prime}$. With $d\, x^\prime = d\,r^\prime
\cos\phi^\prime - r^\prime \sin\phi^\prime\,d\phi^\prime$, $
d\,y^\prime = d\,r^\prime \sin\phi^\prime + r^\prime \cos\phip \,d\,
\phip$,  $i_{\pd \phip}\,d\,x^\prime= -\rp\sin\phip$,
 $i_{\pd
\phip}\,d\,y^\prime= \rp\cos\phip$, $i_{\pd \phip}\,d\,z^\prime= 0$
 \BE{ i_{\widetilde{d\phi}(Y)}\gm 1\vert_{\rp=a,z^\prime=0}=
\frac{1}{a R}\{ d\,r \,\sin(\phi-\phip) + r
\,d\,\phi\,\cos(\phi-\phip)  \}
 }
Further with $$\HASHY 1=\rp\,d\,\rp\wedge d\,z\wedge d\,\phip =
d\,f_1\wedge d\,f_2\wedge \LERAYY 1$$
$$\LERAYY 1=a\,d\,\phip$$  and one has \footnote{The term in  the integrand proportional to
 $ \sin (\phi-\phip) $ integrates to zero.}
\BE{{\cal A}_X(r,\phi,z)= r\,d\,\phi \frac{1}{4\pi} \int_{0}^{2\pi}\frac{\lambda
 \cos(\phi-\phip)  }{(r^2+a^2 +z^2 -2ar\cos(\phi-\phip))^{1/2}}   \,d\,\phip }
or with $\Psi=\phip-\phi$ and noting that the integrand is an even
periodic function of $\Psi$: \BE{{\cal
A}_X(r,\phi,z)=-r\,d\,\phi\int_0^{2\pi}
\frac{\,\lambda\cos\Psi\,d\,\Psi} {(r^2+a^2 +z^2 -2ar\cos\Psi)^{1/2}}
} with magnitude  independent of $\phi$. For a constant current $\breve I$
in the circular loop the constant  $\lambda= 4\pi\,\mu_0\, \breve I a$.
 In MKS units the physical unit for   $  \TYPE 2 J {} $
and hence $\TYPE 1 {\bf h} {}$ is the ampere and that for $\JJJ$ is
$\mu_0$ ampere/m. The methodology here illustrated for a circular planar coil is directly applicable to any open or closed current carrying conductor of arbitrary shape in space and possibly composed of piecewise smooth connected segments.

\newcommand\xii{\TYPE {0} {\hat\xi} {}}

To verify that when  $\bfA^D[{\hat\psi}] = \int_X {\cal A}\wedge
\HASH {\hat\psi}$ the distribution $\bfA^D$ lies in the gauge
satisfying $\delta \bfA^D=0$ one must compute \BAE{
\delta\bfA^D[\xii]&=\bfA^D[d\xii]\\&= \int_X {\cal A} \wedge \HASHX
\,d\xii=\int_X d\xii\wedge\HASHX {\cal A}\\&=-\int_X \xii d \HASHX
{\cal A}=-\int_X \xii \,\,\HASHX \delta {\cal
A}\\&=-\lambda\int_X\xii \HASHX\int_{\SIG 1(Y)}
i_{\widetilde{d\phi(Y)}} \delta_X \gm 1\,\LERAYY 1 }

But $$\delta_X\gm 1 = d_Y\gm 0   $$
$$i_{\widetilde{d\phi(Y)}}\,d_Y\gm 0 = \frac{\partial}{\partial\phi^{\prime}}\,\gm 0  $$
so $$ \delta \bfA^D[\xii]=0$$ since $\LERAYY 1=a\,d\,\phi^{\prime}$
and $\SIG 1(Y)$ is a closed curve. This is physically equivalent to
the statement that the current in the coil  is \lq\lq
conserved\rq\rq.

One can just as easily work in a gauge invariant manner by solving
the equation
\BE{\Delta \poph=\delta \bbfJ \label{heqnn} }
All solutions to (\ref{mag1}) and (\ref{mag2}) must satisfy (\ref{heqnn}).
As noted above (\ref{aaa}),  for a smooth source $\delta J$,
the solution in free space is \BE{\TYPE 1 {\poph} {}(Y)= \int_X \gm 1
\wedge \HASHX (\delta \bbfJ)_X.\label{aa} } For a distributional source
the solution
\BE{\mu_0 {\TYPE 1 h {}}^D[\hat\psi]= \TYPE 1 {{\cal
J}^D} {}  [\TTilde \Gamma \hat\psi] } (cf (\ref{step})) is modeled on the
representation following directly from (\ref{aa}):

\BE{ \mu_0 {\TYPE 1 h {}}^D [\hat\psi]=\int_X\int_Y {\cal J}_Y
\wedge \HASHY {\gmp 1} \wedge \HASHX \hat\psi_X } In these equations
the operator $\TTilde\Gamma$ is defined as $\Gamma$ but with $\gm 1$
replaced by ${\gmp 1}$ where
 \BE{{\gmp 1}=\delta_Y\HASHY \gm 1.} Thus for the distributional source $$
{\TYPE 1 {\cal J} {}}^D=\lambda \widetilde W \delta_{\SIG 1}
$$ for some vector field $W$ on $\bf R^3$ one has

\BE{ \mu_0 {\TYPE 1 h {}}^D[\TYPE 1 {\hat\psi} {}] = \int_X
 \TYPE 1 {{\cal H}_X} {}\,\wedge
 \HASHX \hat\psi_X } where \BE{ {{\cal H}_X}=\int_{\SIG
1}\lambda i_{W_Y} {\gmp 1 \,\LERAYY 1} } Following (\ref{lambda})
the choice of $W$ and $\SIG 1$ determines the physical dimensions of
$\lambda$. \footnote{For a loop with a non-constant but steady
current, $\widetilde W= d\phi$ and $\SIG 1=a\,d\phi$ one could take
$\lambda =4\pi \mu_0 I_0 a f(\phi)$ for some physically dimensionless
function $f$ of $\phi$ to describe a coil with varying resistivity.}

Although (\ref{aa}) solves (\ref{heqnn}), in order  to qualify as a magnetostatic field one must verify that it satisfies
$\delta \poph =0$ and $d\,\poph={\bbfJ}$. Now for some $\DD\in {\bf R}^3$  containing the smooth source ${\bbfJ}$ with
$$h_Y=\left(\gm 1\,,  {{\bbfj}}_X\right)_{\DB}$$
one has
$$(\delta h)_Y= \left(   \delta_Y \gm 1\,,   {{\bbfj}}_X   \right)_{\DB}
 = \left(   d_X \gm 0\,,   {{\bbfj}}_X   \right)_{\DB} $$

$$ =\BRK {\gm 0} {\delta_X\,{{{\bbfj}}_X}} + \int_{\partial\DB} \gm 0 \wedge \HASH_X {{\bbfj}}_X$$

$$ =\BRK {\gm 0} {\HASH_X\, d\,{{{\bbfJ}}_X}} + \int_{\partial\DB} \gm 0 \wedge  {{\bbfJ}}_X$$
Hence with $d{{\bbfJ}}=0$, and ${{\bbfJ}}\vert_{S_\infty}=0$ and (by explicit calculation)
  $$\TYPE {\epsilon\to 0} {Lim}  {} \int_{S_\epsilon} \gm 0 \wedge
 {{\bbfJ}}_X=0$$ one has
  \BAE{\delta \poph=0\label{OK}}
Writing  $\phi=\poph$ and $\psi=\hat\psi$ in (\ref{key3})
$$ \BRK{\delta {{\bbfJ}}}{\hat \psi} - \BRK{d\poph}{d\hat \psi} - \BRK{\delta \poph}{\delta\hat
\psi}=\int_{\partial\DB}(\delta\,\poph\,\wedge \HASH\hat\psi
-\hat\psi\,\wedge \HASH d\,\poph)$$
But $\delta \poph=0$ and $\BRK{\delta {{\bbfJ}}}{\hat\psi}=\BRK{{{\bbfJ}}}{d\hat\psi}$.
It remains to calculate $\int_{\partial\DB}\HASH\,d\poph\,\wedge
\hat\psi$ with $\poph=\BRK{\gm 1}{\delta {{\bbfJ}}}$ and so
$(d\poph)_Y=\BRK{\delta_X\gm 2}{\delta_X {{\bbfJ}}_X}$. Thus
$$ \int_{(\partial B_\epsilon)_Y} (\HASH d\poph)_Y\,\wedge \hat\psi_Y=  \int_{(\partial
B_\epsilon)_Y}\BRK{\HASH_Y\delta_X\gm 2}{\delta_X {{\bbfJ}}_X} \,\wedge
\hat\psi_Y$$
$$=\BRK{\delta_X \Lambda_X} {\delta_X  {{\bbfJ}}_X}$$ where $\Lambda_X= \int_{(\partial
B_\epsilon)_Y}\HASH_Y\gm 2 \,\wedge \hat\psi_Y$. In the limit as
$\epsilon\to 0$ one has by explicit computation that this integral
is zero, hence $$ \BRK {d\poph}{d\hat\psi}=\BRK{{{\bbfJ}}}{d\hat\psi}$$ But
since $d\,\hat\psi $ is an arbitrary test form it follows that
 \BAE{d\poph={{\bbfJ}}\qquad on\,\,\,\DD\label{OKK}}
 These arguments generalize immediately to the singular distributional case.

\section{Magnetostatic Fields from a Steady Helical Line Current}
\label{chhelix}
The derivation above of the potential for the magnetic field due to a circular line current was somewhat labored. However no further steps are required to find the potentials due to more complex geometries once one parametrises the source in terms of the geometry of its support in space.

Suppose a helical wire with pitch $p>0$ and radius $a$ is the space curve $(r=a, \phi=\frac{\sqrt{1-p^2}}{a}\sigma, z=p\,\sigma)$ in cylindrical polar coordinates. The parameter $\sigma$ is the arc-length parameter  so $0<\sigma<L$
for a helix of  length $L$. It is convenient to introduce $P>0$ with $a^2\,P^2+p^2=1$. Then the unit tangent to the helix is the direction $W=p\frac{\partial}{\partial z} + a\,P\frac{\partial}{\partial\phi}$. If one chooses to describe the helix in terms of the foliating set ${\bf f}=\{f_1=r-a, f_2= z-\frac{p}{P}\,\phi   \}$ this yields a Leray form $\Omega_{\bf f}=-r\,d\,\phi$. Then, in this case, the source  $ \JJJ^D=\lambda\,\widetilde W  \,\DIRAC 1 $ with support on the helical wire yields the magnetostatic potential $1-$form
\BE{
{\cal A} (r,\phi,z) = {\cal A}_r (r,\phi,z)\,d\,r +  {\cal A}_\phi (r,\phi,z)\,r\,d\,\phi +  {\cal A}_z (r,\phi,z)\,d\,z \label{helixsol}
}
where
\BE{{\cal A}_r= \frac{\lambda}{4\pi}\int_0^{L} \frac{d\,\sigma}{R}\,\sin{(P\sigma-\phi)}
}
\BE{{\cal A}_\phi= -\frac{\lambda\, a^2\,P^2}{4\pi}\int_0^{L} \frac{d\,\sigma}{R}\,\cos{(P\sigma-\phi)}
}
\BE{{\cal A}_z= -\frac{\lambda\, a\,p\,P}{4\pi}\int_0^{L} \frac{d\,\sigma}{R}
}
with
\BE{ R^2=r^2+a^2+z^2+\sigma^2\,p^2 -2\,z\,p\,\sigma-2a\,r\,\cos(P\sigma-\phi)
}
For a constant current  $\breve I_0$ in the helix the constant $\lambda=4\pi\,\mu_0 \breve I_0 a$.
The potential ${\cal A}$ yields a global description of the field in terms of elliptic integrals that depend on the geometry of the helix specified by the parameters $L,a$ and $p$ or $P$. Such parameters offer natural scales that are useful for defining dimensionless variables that in turn can be used to generate multipole or asymptotic expansions of the above integrals.    The practical generation of high intensity uniform magnetic fields by carefully designing coils with complex helical windings is of paramount importance in the construction of undulators and free electron lasers.

\section{Magnetostatic Fields from a Steady Helical Surface Current}
\label{chhelix}

It is natural to model a solenoid composed of {\it closely wound current-carrying turns} by a  surface current source. Such a current, regarded as a vector field on the solenoid surface $\Sigma^2_{\bf f}$, can have an arbitrary direction $W\vert_{\Sigma^2_{\bf f}}$ and the distributions $\JJJ=\kappa_0\,\widetilde W \delta_{\Sigma^2_{\bf f}}$ are well suited to model such a source.  Suppose $\Sigma^2_{\bf f}$ is a right circular cylinder of radius $a$ and length $L_0$ and the surface current is \lq\lq painted" on it with helical strokes of pitch $p=\sqrt{1-a^2\,P^2}$. Thus the integral curves of  $W=p\frac{\partial}{\partial z} + a\,P\frac{\partial}{\partial\phi}$ are each similar to the helix above. To construct $\delta_{\Sigma^2_{\bf f}}$ one parametrises the solenoid surface as $\{r=a,z=\rho,\phi=\sigma\}$ in cylindrical polars with $0<\sigma<2\pi$ and $0<\rho<L_0$ and takes ${\bf f}=r-a$ to generate $\Omega_{\bf f}=r\,d\,\phi\wedge d\,z$. It follows that, in this
case, the source  $ \JJJ^D=\kappa_0\,\widetilde W  \,\DIRAC 2 $ with support on the cylindrical surface yields the magnetostatic potential
\BE{
{\cal A} (r,\phi,z) = {\cal A}_r (r,\phi,z)\,d\,r +  {\cal A}_\phi (r,\phi,z)\,r\,d\,\phi +  {\cal A}_z (r,\phi,z)\,d\,z \label{helixsol1}
}
where
\BE{
{\cal A}_r = a^2 P \frac{\kappa_0}{4\pi}\int_0^{2\pi} \left( \int_0^{L_0} \frac{d\,\rho}{R}\sin(\sigma-\phi) \right)\,d\,\sigma
}
\BE{
{\cal A}_z = -a p \,\frac{\kappa_0}{4\pi}\int_0^{2\pi} \left( \int_0^{L_0} \frac{d\,\rho}{R} \right)\,d\,\sigma
}
\BE{
{\cal A}_\phi = -a^2 P \frac{\kappa_0}{4\pi}\int_0^{2\pi} \left( \int_0^{L_0} \frac{d\,\rho}{R}\cos(\sigma-\phi) \right)\,d\,\sigma
}
with $$ R^2=r^2+a^2+z^2+\rho^2-2z\rho -2ar\cos(\sigma-\phi)$$
The double integrals above can be reduced to quadratures involving Elliptic integrals.

\section{Electrostatics with  a Singular Distributional Charge Source in a Domain with a Boundary}
\label{ch7}

Returning to the electrostatic example discussed in  section [\ref{ch3}] in Cartesian coordinates, suppose a uniform charged
straight wire is inserted in the direction of the $y$ axis,  between  the grounded ($\widetilde\phi=0$)
planes at $z=0$ and $z=L$,  at a position with $z=z_0\,\, (0<z_0<L)$ and $ x=0$.
Then with ${\bf f}=\{x,z-z_0\}$ one has a  $0-$form {\it 0-layer }
distributional source \BE{\bar\rho^D=\lambda \DIRAC 1} and
\BE{\phi^D[\hp]=\bar\rho^D[\GGG\hp]= \int_\DD {\cal P}_X \HASHX \hp_X}
 with \BE{
{\cal P}_X=\lambda \int_{\SIG 1} \GGG_{XY}\,
 \LERAY 1
 }
 Since $\LERAYY 1 =d\yp$ one finds from (\ref{GSUM})
 \BE{ {\cal P}_X(x,y,z)= \frac{\lambda}{\pi L} \sum_n \sin\frac{n\pi z_0}{L}  \sin\frac{n\pi
z}{L}  \int_{-\infty}^{\infty} \frac{dk_x}{k_x^2 +
\frac{n^2\pi^2}{L}}  e^{ik_x x}} The $k_x$ integration can be done
by contour integration in the complex $k_x$ plane. One has \BE{
\begin{cases}
\int_{-\infty}^{\infty} \frac{\,e^{isx}\,ds}{s^2 +
b^2}=\frac{\pi}{b}e^{-bx} & \qquad\qquad  xb>0\\{}\\
\int_{-\infty}^{\infty} \frac{\,e^{isx}\,ds}{s^2 +
b^2}=\frac{\pi}{b}e^{bx} & \qquad\qquad  xb< 0\\
\end{cases}
} yielding \BE{ {\cal P}_X(x,y,z)=
\begin{cases}
 \frac{\lambda}{\pi}\sum_{n}
\frac{1}{n}\sin\frac{n\pi z_0}{L}  \sin\frac{n\pi z}{L}\, e^{-
\frac{n\pi\,x}{L}} & \qquad\qquad x>0\\{}\\
 \frac{\lambda}{\pi}\sum_{n}
\frac{1}{n}\sin\frac{n\pi z_0}{L}  \sin\frac{n\pi z}{L}\, e^{
\frac{n\pi\,x}{L}} & \qquad\qquad x<0
\end{cases}
}
This agrees with the computation in \cite{ams}

\section{Time Dependent Electromagnetic fields and Smooth Sources}
\label{ch8}

\newcommand\JP{{\bar{\cal J}}}

\newcommand\bfG{{{\mathbf \Gamma}_\omega}}

\newcommand\bfGR{{\mathbf \Gamma}_R}

\def\FF{{\cal F}}

\def\FFS{{\cal F}^{\star}}

\def\tXY{{t-\frac{\vert {\bf x}-{\bf y} \vert}{c}}}

\def\tXY{{t-\frac{\vert {\bf x}-{\bf y} \vert}{c}}}

\def\tp{{t}^\prime}

\def\RR{{\vert {\bf x}-{\bf y} \vert}}
\def\bfv{{\bf v}}
\def\Rc{\frac{\RR}{c}}

When the smooth electromagnetic  sources $\rho$ and ${{\bbfJ}}$ depend on $t$ they may generate time dependent electromagnetic fields that must satisfy (\ref{M11}), (\ref{M12}), (\ref{M13}), (\ref{M14}). These equations offer a well-posed initial-boundary value problem that is traditionally approached by finding time-dependent potentials $\phi$  and $\bfA$ that satisfy the source-driven wave-equations (\ref{MAX1}), (\ref{MAX2}) (subject to initial and boundary conditions) in the Lorentz gauge. When the sources are distributional one seeks distributional potentials subject to similar conditions. The distributional formulation given above for real static field configurations generalizes  without difficulty to the time dependent situation. The essential modification is to define functionals on the space of $t$ dependent complex test forms on ${\bf R^3}$  and exploit the Fourier transform  \cite{taylor} of such forms. Thus if $\hat\psi$ is a $t$-dependent test $p$-form on $X={\bf R^3}$ its Fourier transform is the $\omega$-dependent test $p$-form   $\FF\hat\psi$ on $X={\bf R^3}$ defined by
\BE{(\FF\hat\psi)(X,\omega)= \frac{1}{\sqrt{2\pi}}\int_R \hat\psi(X,t)\,\exp(i\omega t)\,d\,t
}
with inverse
\BE{\hat\psi(X,t)= \frac{1}{\sqrt{2\pi}}\int_R (\FF\hat\psi)(X,\omega)\,\exp(-i\omega t)\,d\,\omega
\label{F1}}
The fundamental  $t$-dependent $(p,p)$ type double-form  is the real part of the Fourier transform of  \footnote{Note that the phases in (\ref{F1}) and (\ref{popp}) are chosen so that for $0<r<\infty,\quad t>0$ $\exp(\pm i \frac{\omega}{c} r). \exp(\mp i\omega t)=\exp(\pm i\frac{\omega}{c}(r-ct) )$ describes a radially outgoing wave.}
\BE{\KK p= \exp\left({-i\,\frac{\omega}{c}{\vert {\bf x}-{\bf y}\vert}}\right)\,\,\gm p.\label{popp}}

It is useful, then, to define the singular complex-linear integral operator $\bfG$ on all $p-$forms $\psi$ by:
\BE{
(\bfG\psi)(Y,\omega)= \int_X \KK p \wedge \HASHX \psi(X,\omega)
}
Suppose that  $\JP$ is a smooth $t-$dependent $p-$form source (regular at spatial infinity) that enters into the equation
\BE{
\Delta_X C(X,t) + \frac{1}{c^2}\,\, \ddot C(X,t) =\JP(X,t)\label{Ceqn}
}
for the time-dependent $p-$form $C$ on $X= {\bf R^3}$. Then

\BE{ (\FF C)(X,\omega)= ( \Delta_X - \frac{\omega^2}{c^2} )^{-1}\, (\FF \JP )(X,\omega)
}
Since the singularity structure of $\KK p$ is the same as that for $\gm p$ it is straightforward to show that  a particular solution, regular at spatial infinity, is given by
\BE{
(\FF C)(X,\omega)= \int_Y \KK p \wedge \HASHY \, (\FF \JP) (Y,\omega)
\label{pop1}}
 For each $\omega$ the function $\FF C$ may be associated with a functional by applying the operation $\int_X \HASHX \hat\alpha(X,\omega)$ to define:
\BE{
(\FF C)[\hat\alpha](\omega)=\int_X (\FF C) (X,\omega) \wedge  \HASHX \hat\alpha (X,\omega)
}
But from (\ref{pop1}) this may be expressed
\BE{
(\FF C)[\hat\alpha](\omega)=
\int_Y(\FF \JP)(Y,\omega) \wedge \HASHY \, (\bfG \hat\alpha)(Y,\omega)\equiv (\FF\JP)^D[\bfG \hat\alpha](\omega)
}
Thus from (\ref{popp}) and  Fourier inversion
\BE{
C(X,t)=\int_Y\,\gm p \wedge \HASHY \, \JP\left(Y, \tXY\right)\label{motivate}
}
and
\BE{
C[\hat\alpha](t)= \frac{1}{\sqrt{2\pi}} \int_R d\,\tp \int_Y  \JP (Y,\tp)  \wedge \HASHY \int_X \gm p \wedge \HASHX (\FF\hat\alpha)\left(X,t-\tp -\Rc\right)
}
or
\BE{
C[\hat\alpha](t)= \frac{1}{\sqrt{2\pi}} \int_R d\,\tp \int_Y  \JP \left(Y,t-\tp-\Rc\right)  \wedge \HASHY \int_X \gm p \wedge \HASHX (\FF\hat\alpha)\left(X,\tp\right)
}

Motivated by (\ref{motivate}) the distribution $C^D$ associated with the smooth $p-$form $C(X,t)$ is defined by
\BE{
C^D[\hat\beta]=\int_R d\,t \int_X C(X,t) \wedge \HASHX \,\hat\beta(X,t)
}
Hence
\BE{C^D[\hat\beta]=\int_R d\,t \int_X \int_Y \gm p \wedge \HASHY \JP\left( Y,t-\Rc\right) \wedge \HASHX \hat\beta(X,y)
}
or with $t\mapsto \tp-t-\Rc$ and interchange of $X$ and $Y$
\BE{C^D[\hat\beta]=\int_R d\,\tp \int_X (\bfG_R\hat\beta)(X,\tp) \wedge \HASHX \,\JP (X,\tp)}
where
$$(\bfGR\hat\beta)(X,\tp)\equiv \int_Y\gm p \wedge \HASHY \hat\beta\left(Y,\tp + \Rc\right)
$$
Thus finally
\BE{C^D[\hat\beta]=\int_Rd\,\tp \JP(X,\tp) \wedge \HASHX (\bfGR\hat\beta)(X,\tp)\equiv \JP^D[\bfGR\hat\beta]\label{pop3}}
where $\JP^D$ is the distribution associated with the {\it smooth} time-dependent $p-$form $\JP$. If one now contemplates a distributional source $\JP^D$ that is not associated with a smooth time-dependent $p-$form on ${\bf R}^3$ then (\ref{pop3}) offers a particular  distributional solution  $C^D$ to (\ref{Ceqn}) based on a retarded extension of the fundamental solution $\gm p$.

The above solution is immediately applicable to the problem of finding the retarded electromagnetic potentials $\bfA$ and $\phi$ in free space, where $c^2\,\epsilon\mu=1$, for {\it any} smooth time dependent source.
Thus from section (\ref{ch time}) one sees that with  a smooth source $\JP = -\frac{\rho}{\epsilon}$ in equation (\ref{Ceqn}) the smooth $0-$form solution $\phi$  associated with $C^D$ describes  a free-space scalar potential solution to (\ref{MAX2}) and that  with  a smooth source $\JP = \mu\HASH {{\bbfJ}}$ in equation (\ref{Ceqn}) the smooth $1-$form solution $\bfA$ associated with $C^D$ describes  a free-space vector potential solution to (\ref{MAX1}).

In the absence of losses due to conduction the {\it convective current} spacetime $3-$form is given as $j=\rho_0\star\widetilde V$ where the unit future-pointing  time-like vector field $V$ convects {\it proper-charge density} $\rho_0$. In a local spacetime co-basis $\{c\,d\,t,e^1,e^2,e^3\}$ adapted to $U$ the source velocity
 $\widetilde V=\gamma\left(-c\,d\,t+ \frac{\widetilde\bfv}{c}\right)$ where
 $\gamma^{-2}=1- \frac{\vert \bfv \vert^2}{c^2}$
 and $\widetilde\bfv=\sum_{k=1}^3 v_k\,e^k$
is the instantaneous time-dependent (Newtonian) $3-$velocity $1-$form  in ${\bf R}^3$.
  Since
\BE{
j=-{{\bbfJ}}\wedge d\,t + \rho \hash 1=\rho_0\star \widetilde V} one finds immediately
$$ \HASH {{\bbfJ}}= -\rho \widetilde\bfv $$
The analysis above  also offers distributional solutions to (\ref{MAX1}) and (\ref{MAX2}) in terms of scalar and $1-$form distributional sources. The electromagnetic field associated with an arbitrarily moving charge point source is space can now be modeled in terms of a moving Dirac distribution in space.

\section{Time Dependent Electromagnetic fields and Distributional Sources:
Moving Dirac Distributions}
\label{ch9}

The above section formulates a distributional solution to (\ref{Ceqn}) associated with a  $t-$ dependent $p$-form distributional source  $\JP^D$ on ${\bf R}^3$.  $0-$layer singular distributions associated with electromagnetic sources constrained to moving curves or surfaces in space can be constructed in terms of a family of distributions $\DIRACC {3-k}$ defined for $k=1,2$ by
\BE{\DIRACC {3-k}[\TYPE {0} {\hat\beta} {} ]=\int_R\,\,\ \DSIG {3-k}[\TYPE {0} {\hat\beta} {}]   \,\,d\,t
}
Then
\BE{\widetilde W\,\DIRACC {3-k}[\TYPE {1} {\hat\beta} {} ]=\int_R\,\,\ \widetilde W(t)\,\DSIG {3-k}[\TYPE {1} {\hat\beta} {}]   \,\,d\,t
}
for some $t-$dependent $1-$form $\widetilde W$ in ${\bf R}^3$ with support on  $ \Sigma_{\bf f}^{3-k} $.

 For the case $k=3$  the support of $\DSIG {3-k}$ is a moving point $p_0(t)$ in $X={\bf R}^3 $. Then the  worldline history of the moving point is encoded into the support of a distribution $\DIR$ where
 \BE{\DIR[ \hat\beta]= \int_R \hat\beta(p_0(t),t)\,d\,t
 }
 If the history of $p_0\in X$  is parameterized by  $t$ it is the  spacetime curve
\BE{
\tp\mapsto ({\bf x},t)=(\XX \tp,\tp)
}
Then
\BE{\DIR[\TYPE {0} {\hat\beta} {} ]=\int_R\,\,\,\,\TYPE {0} {\hat\beta} {}   \left(\XX \tp, \tp   \right)   \,\,d\,\tp
}
and
\BE{\widetilde W\,\DIR[\TYPE {1} {\hat\beta} {} ]=\int_R\,\,\,\TYPE {0} {(i_{W(\tp)}\,\hat\beta)} {}   \left(\XX \tp, \tp   \right)   \,\,d\,\tp
}

One may now easily calculate \BAE{
  \DIR [\bfGR \hat\beta] =&\int_R (\bfGR\hat\beta ) ( \XX \tp,\tp )\,d\,\tp \\&
  =\int_R\,d\,\tp \int_Y \TYPE {0} {\gamma} {}  (\XX \tp, Y   )\,\, \HASHY \hat\beta \left( Y, t+ \frac{\vert \XX \tp -{\bf y}  \vert }  {c}  \right)
}
Changing variable $\tp\mapsto t=\tp + \frac{\vert \XX \tp -{\bf y}  \vert }  {c}$ at fixed $Y$  with Jacobian $\JAC$ defined by
$$d\,t=\JAC^{-1}(Y, \tp  )\,d\,\tp$$ yields
\BE{
 \DIR [\bfGR \hat\beta] =\int_R d\,t \,\int_Y \left( \JAC(Y,\tp) \TYPE {0} {\gamma} {} ( \XX \tp ,Y )  \right) \,\,\HASHY \hat\beta(Y,t)
}
where $\tp \equiv \hat\tp (Y,t)$ solves the equation $ \tp=t - \frac{\vert \XX \tp -{\bf y} \vert }  {c}   $.
Thus \BE{ {\cal Z}(Y,t)= \JAC(Y,\tp)   \,\, \TYPE {0} {\gamma} {} {(\XX \tp, Y   )}
}
is the $0-$form associated with the distributional source  $\DIR \, [\bfGR \hat\beta]$ at the field point $Y$ at the instant $t$.
It is straightforward to calculate the Jacobian in terms of $\XX t$ and its derivative $\XXP t$:
\BE{ \JAC^{-1}(Y,t)= \left( 1+ \frac{1}{c} \frac{(\XX \tp - {\bf y})\cdot (\XXP \tp)}  {\vert \XX \tp - {\bf y}   \vert    }   \right)
}
The instantaneous Newtonian $3-$velocity of the point support is ${\bf v}(t)\equiv \XXP t$ and if one introduces the  Euclidean unit vector
$${\bf n}(Y,t)\equiv \frac{ {\bf y} - \XX \tp  } {   \vert {\bf y} - \XX \tp \vert }$$ connecting the field point ${\bf y}$ at time $t$ to the source point  $\XX \tp $at the earlier time $\tp=\hat\tp (Y,t)$,   then the Jacobian inverse takes the form
\BE{
\JAC^{-1}(Y,t)= \left(  1- g\left(\frac{{\bf v}(\tp)}{c}   ,  {\bf n}(Y,t) \right)  \right)
}
where $\tp \equiv \hat\tp (Y,t)$ solves the equation $ \tp=t - \frac{\vert \XX \tp -{\bf y}  \vert }  {c}   $ and $g$ denotes the Euclidean metric tensor on ${\bf R}^3$

By contrast to the smooth electromagnetic sources $\JP$ discussed in section (\ref{ch8}) one models a moving point source (with electric charge $q$ and $3-$velocity ${\bf v}(t)$)  for the scalar potential $\phi^D$ by
\BE{\TYPE {0} {\JP^D} {}=-\frac{q}{\epsilon}\, \DIR} and for the  potential $1-$form $\bfA^D$ by
\BE{\TYPE {1} {\JP^D} {}=-q\mu\,{\widetilde{\bf v}}(t)\, \DIR
}
It then follows immediately from (\ref{pop3}) that the distribution $\phi^D$ may be associated with
\BE{
\TYPE 0 {\cal P} {}(Y,t) = -\frac{q}{\epsilon}\, {\cal Q}(Y,t) \, \TYPE {0,0} {\gamma} {} ({\xx}(\tp),Y)
}
while the distribution $\bfA^D$ may be associated with
\BE{ \TYPE 1 {\cal A} {} (Y,t)=-q\,\mu\,{\cal Q}(Y,t)\, \, i_{{\bf v}(\tp)}\,\TYPE {1,1} \gamma {} ( {\xx}(\tp),Y  )
}
where $\tp=\hat\tp(Y,t)$ as above.
These are the classic  Lienard-Weichert    potentials for a moving point charge \cite{jackson}.
It is of interest to verify from $\HASH {{\bbfJ}}^D=-\widetilde {\bf v}\rho^D$ that the distributions ${\bf A}^D$ and $\phi^D$ satisfy the gauge condition  $\delta\,{\bf  A}^D + \epsilon\mu \dot\phi^D$=0.
It may also be noted that  by using the appropriate fundamental double-forms in place of $\TYPE {0,0} {\gamma} {}$ and $ \TYPE {1,1} {\gamma} {}$ these potentials maintain their structure for solutions in ${\cal D}\subset Y $ that satisfy appropriate boundary conditions on ${\partial D}$.


\section{Conclusions}
\label{ch10}

 With the aid of properties of the fundamental double-form of bi-degree $(p,p)$
 associated with the Hodge-de Rham operator $\Delta$ on differential forms,
 a distributional framework for analysing equations of the form $$\Delta\TYPE p \Phi {} + \lambda^2\TYPE p \Phi {}=\TYPE p {\cal S} {}$$ on ${\bf R}^3$ has been established. A set of  $r-$layer Dirac singular distributions  with supports on (moving) embeddings in ${\bf R}^3$ has been constructed and finds application in electromagnetic source modeling. The framework has been illustrated by explicitly calculating the fields associated with a current carrying circular and helical coil, a finite length solenoid with a helical surface current density, a uniformly charged wire between  two conducting plates and an arbitrarily moving point charge in free space.
 With the aid of  $r-$layer Dirac singular distributions  many fields with more complex distributional sources  can be  readily reduced to quadratures once one parametrises such sources in terms of the geometry of their support in space.

\section*{Acknowledgment}
The author is grateful to colleagues at the Cockcroft Institute
for valuable
discussions and to the EPSRC  for a Springboard Fellowship and financial
support for this research which is part of the Alpha-X
collaboration.



\end{document}